# Research on Modeling Units of Transformer Transducer for Mandarin Speech Recognition

*Li Fu*[1,†], *Xiaoxiao Li*[1,†], *Libo Zi*[1,†]

[1]Jingdong Digits Technology Holding Co., Ltd. (JD Digits), Beijing, China
{fuli3, lixiaoxiao10, zilibo}@jd.com

## Abstract

Modeling unit and model architecture are two key factors of Recurrent Neural Network Transducer (RNN-T) in end-to-end speech recognition. To improve the performance of RNN-T for Mandarin speech recognition task, a novel transformer transducer with the combination architecture of self-attention transformer and RNN is proposed. And then the choice of different modeling units for transformer transducer is explored. In addition, we present a new mix-bandwidth training method to obtain a general model that is able to accurately recognize Mandarin speech with different sampling rates simultaneously. All of our experiments are conducted on about 12,000 hours of Mandarin speech with sampling rate in 8kHz and 16kHz. Experimental results show that Mandarin transformer transducer using syllable with tone achieves the best performance. It yields an average of 14.4% and 44.1% relative Word Error Rate (WER) reduction when compared with the models using syllable initial/final with tone and Chinese character, respectively. Also, it outperforms the model based on syllable initial/final with tone with an average of 13.5% relative Character Error Rate (CER) reduction.

## 1. Introduction

With the rapid development of Deep Neural Network (DNN), end-to-end DNN systems have become increasingly popular in speech recognition due to the advantages of easiness in training process and model modification for a new language with little linguistic knowledge. Among these end-to-end Automatic Speech Recognition (ASR) frameworks, Recurrent Neural Network Transducer (RNN-T) is proposed to integrate acoustic and linguistic information during a speech recognition task, which provides promising potential on accuracy and efficiency [1]. Moreover, since RNN-T is trained to learn the alignments between the acoustic feature and the label feature in a frame-synchronous manner, it is naturally adapted to streaming ASR applications [2, 3].

Mandarin, as one of the most spoken languages in the world, is an important object for ASR. Although several optimized techniques have been explored, there is still room for improvement in Mandarin RNN-T to meet the requirements in various practical scenarios. RNN-T was first proposed for English speech recognition [4]. The model architecture was mainly composed of two RNNs, which were used for acoustic feature encoding and label feature encoding, respectively. These two encodings were jointed to learn the alignments between each acoustic frame and each label token. Inspired by the work of [4], Wang et al. explored RNN-T for a Chinese large vocabulary speech recognition task using Chinses character as modeling unit [5]. Also, convolutional layers were added at the beginning of the acoustic encoding network to improve the convergence performance. More recently, transformer or self-attention based end-to-end ASR models have been proven very effective [6, 7]. To improve the performance of RNN-T in accuracy and parallel-in-time computation, self-attention transducer was proposed by replacing the RNN architecture in RNN-T with multi-head self-attention mechanism [8]. In [9], a convolution and transformer based RNN-T was proposed for Chinese speech recognition and Minimum Bayes Risk (MBR) was applied to improve the model training.

Besides model architecture, modeling unit is one of the main factors affect the performance of RNN-T [10]. Using different modeling units, the number of modeling units that the model needs to classify and the length of label features that the model needs to extract will be very different. Most of the existing Mandarin RNN-T works focus on the architecture modification, while pay relative less attention on the modeling units. The related works about modeling units for Mandarin ASR are studies in some other end-to-end frameworks. Zhou et al. compared different modeling units on Mandarin ASR tasks by end-to-end attention-based model with transformer on HKUST dataset, and Chinese character based model achieved the best result [11]. With large vocabulary speech data, two end-to-end ASR frameworks based on Connectionist Temporal Classification (CTC) and attention were compared with various modeling units. The results showed that all modeling units achieved approximate performance in the CTC model, while Chinese character performed the best in the attention-based method [12].

To improve the performance of RNN-T for Mandarin speech recognition, we research on the effect of different modeling units for transformer transducer. First, instead of using pure transformer encodings, a combination of transformer and RNN architecture is proposed for Mandarin speech recognition. As referred in [13], label feature encoding of RNNs still gives better performance than the transformer and is smaller in size. Thus, we construct our streamable transformer transducer using truncated transformer blocks [14] for acoustic feature encoding with convolutional layers at the beginning, and using RNNs for label feature encoding. Second, to improve the performance of our model in various practical scenarios, we present a new mix-bandwidth training method for Mandarin transformer transducer that using training audio with different sampling rates and corpuses. Specifically, all of our experiments are conducted on about 12,000 hours of Mandarin speech data with sampling rate in 8kHz and 16kHz, covering internal speech dataset in different business scopes of

---

† co-first authors

JD Digits and public speech dataset, including AISHELL-1 [15], AISHELL-2 [16], THCHS-30 [17], etc. Finally, the performance of Mandarin transformer transducer is compared with three different modeling units, i.e. syllable initial/final with tone, syllable with tone and Chinese character. Evaluating on different test dataset, our experimental results show that the Mandarin transformer transducer using syllable with tone achieves the best performance, when compared with the models based on the two other modeling units. Our main contributions are shown as follows.

1. We introduce a novel transformer transducer for Mandarin speech recognition with the combination architecture of transformer and RNN.

2. We present a new mix-bandwidth training method for transformer transducer to accurately recognize Mandarin speech with different sampling rates simultaneously.

3. Comparing the performance of different modeling units of Mandarin transformer transducer, our work shows that the model based on syllable with tone outperforms the models based on syllable initial/final with tone and Chinese character.

The remainder of this paper is organized as follows. The details of the proposed transformer transducer are presented in section 2. Section 3 shows the different modeling units for Mandarin speech recognition. Section 4 is the experimental results and discussions. Finally, the conclusions are given in Section 5.

## 2. Transformer Transducer

### 2.1. Model Architecture

Unlike the existing works that using pure transformer to replace the two RNN encodings, our model architecture only uses a transformer based model for acoustic feature encoding, while keeps the RNN structure for label feature encoding. Since the use of transformer extracting language features often requires a deep structure and a large amount of parameters [18, 19], we apply an embedding layer followed by Long Short-Term Memory (LSTM) for label feature encoding to reduce the size of the model [20]. The schematic representation of transformer transducer in this work is shown in Figure 1.

Denote an input speech as $X = [x_1, x_2, ..., x_S] \in R^{1 \times S}$, where $S$ is the number of audio sampling points, whose sampling rate is 8kHz or 16kHz. $X$ is input into the mix-bandwidth module to obtain a sequence feature vector with a unified dimensional, denoted as $F = [f_1, f_2, ..., f_T] \in R^{d \times T}$, where $T$ is the length of the audio feature sequence, $f_t$ is the $d$ dimension feature vector, and $t \in [1, T]$. In the front of the transformer blocks, we added a convolutional structure to improve the feature extraction of acoustic input [5, 13]. Denote the output of the acoustic feature encoding network as $P = [p_1, p_2, ..., p_T]$. In terms of the label feature encoding network, the input label tokens corresponding to different modeling units will be different. Denote an input label sequence as $Y = [y_1, y_2, ..., y_U] \in R^{1 \times U}$, where $U$ is the length of the label sequence. For $u \in [1, U]$, the set of each token $y_u$ is composed of all the elements of the chosen modeling unit and a special label $b$, i.e. blank symbol. The label sequence is first input into an embedding layer, and then passed to a LSTM structure. We denote the higher-level representation of label feature as $Q = [q_1, q_2, ..., q_U]$.

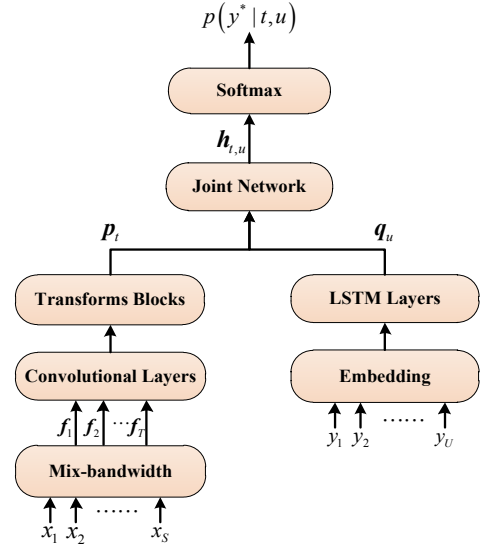

Figure 1: *A schematic representation of transformer transducer in this work.*

Specifically, we use unidirectional LSTM and truncated transformer blocks instead of unlimited transformer. The purpose of this design is to make the model streamable and reduce the latency and computational cost [13]. Then the acoustic encodings and label encodings are passed to a joint network to generate a joint network, which computes the output logits for each input $p_t$ in the acoustic sequence and $q_u$ in the label sequence, yields,

$$z_{t,u} = \psi(W_1 p_t + W_2 q_u + b_1) \quad (1)$$
$$h_{t,u} = W_3 z_{t,u} + b_2 \quad (2)$$

where $W_1$, $W_2$, $W_3$, $b_1$ and $b_2$ are parameters of the joint network model, and $\psi(\bullet)$ is a non-linear function, e.g. tanh.

Finally, the output logits $h_{t,u}$ are then passed to a softmax layer, which defines a probability distribution over the set of output modeling unit and the blank symbol

$$p(y^* | t, u) = \text{softmax}(h_{t,u}) \quad (3)$$

The model can be optimized using Stochastic Gradient Decent (SGD) by computing the required gradients using a dynamic programming algorithm [4].

### 2.2. Mix-bandwidth

Sampling rate is one of the most important attributes of the speech data. Typical ASR models are trained on either 8kHz data or 16kHz data [21]. However, we might obtain a crashing result if we used a ASR model trained on one sampling rate data inferring another sampling rate data. It would be advantageous if a single model could recognize speech with different sampling rates. To achieve this, a mix-bandwidth training method is proposed for transformer transducer. Usually, the speech data features and corpuses are diverse in different sampling rates. Thus, the generalization performance of our model is promising potential for real applications.

### 2.2.1. Feature Calculating

The $d$ dimension Mel-filter bank (fbank) is applied as the input feature, which is orderly calculated by pre-emphasis, framing and windowing, Fourier-transform and power spectrum, and filter banks [22].

Given a 16kHz data, the fbank is calculated with a standard process. Specifically, we use $2N$ point Fourier-transform and obtain $(N+1)$ dimension output in the step of Fourier-transform and power spectrum. Then the final fbank is denoted as $G \in R^{d \times T}$. Unlike 16kHz data, we use $N$ point Fourier-transform for 8kHz data and obtain a $(N/2+1)$ dimension output, and the high frequency is padded with $(N/2)$ zeros to a $(N+1)$ dimension output.

The $d$ dimension fbank is corresponding to the feature from $f_0 = 0kHz$ to $f_h = 8kHz$ (Nyquist Sampling Theorem) in a log-linear manner, yield

$$m_h = 2595\log_{10}(1 + f_h / 700) \quad (4)$$

Then the bandwidth of fbank can be obtain as

$$m_b = m_h / d \quad (5)$$

Finally, the $d$ dimension fbank on a Mel-scale can be easily obtained as referred in [22].

### 2.2.2. Feature Normalization

To improve the generalization of our method, fbank feature is re-scaled to have zero mean and unit variance for each sample data. Since the fbank feature of 8kHz data is padded with zeros in high frequency, a simple normalization will suppress the weight of the audio information itself. To solve this problem, we can calculate the bandwidth of fbank for 8kHz from (4), yields

$$m_l = 2595\log_{10}(1 + f_h / 1400) \quad (6)$$

Then the dimension of fbank for 8kHz can be obtained as

$$n_l = \lceil (m_l - m_b / 2) / m_b + 1 \rceil \quad (7)$$

where $\lceil \bullet \rceil$ is the ceiling operator. For a given $d$ dimension fbank feature, the elements from 0 to $n_l$ correspond the features from 0 to 4kHz, and the elements from $n_l$ to $d$ correspond the features from 4kHz to 8kHz, which is the padding part for 8kHz data. Thus, the normalization result is obtained as

$$G[:n_s,:] = \left[G[:n_s,:] - \mu(G[:n_s,:])\right] / \sigma(G[:n_s,:]) \quad (8)$$

where $n_s = n_l$ and $n_s = d$ for 8kHz data and 16kHz data, respectively, $\mu(\bullet)$ and $\sigma(\bullet)$ are the mean and standard deviation, respectively. We set $d$=80 for feature calculating, and the value of $n_l$ is calculated as 61 for 8kHz data.

The audio feature with the proposed mix-bandwidth is shown in Figure 2. Our experiments show that the model trained with mix-bandwidth improves the performance significantly for both 8kHz and 16kHz audio.

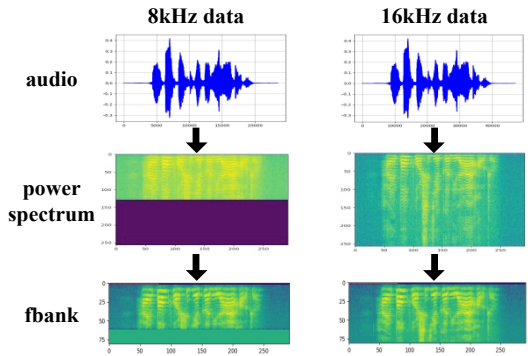

Figure 2: *The audio feature with the proposed mix-bandwidth.*

## 3. Modeling Units

The typical modeling units for Mandarin ASR task are syllable initial/final with tone, syllable with tone and Chinese character [12]. Different from the work in [12], we add the special symbol # into syllable initial/final with tone to align each Chinese character for a Word Error Rate (WER) comparison. Since Chinese character is the basic symbol of Mandarin, it is usually used to label the Mandarin speech dataset. Considering the cost of re-labeling using other modeling units, we use pypinyin library [23] to convert a utterance from Chinese character into the others. An example of converting one Mandarin utterance (*Hello* if translated in English) into different modeling units is shown in Table 1.

Table 1: *Converting a Mandarin example utterance into these different modeling units.*

| Mandarin Utterance | Modeling Units | Converting Results |
|---|---|---|
| 您好 | syllable initial/final with tone | n in2 # h ao3 |
|  | syllable with tone | nin2 hao3 |
|  | Chinese character | 您 好 |

The choice of modeling units is critical to end-to-end Mandarin ASR models. On one hand, the number of classes in different modeling units is quite different. For example, the number of classes of syllable initial/final with tone, syllable with tone and Chinese character in our corpus dataset are 227, 1356, 7228, respectively. Due to homophones in Mandarin, the categories of the two kinds of syllables are much fewer than the category of Chinese character. On the other hand, the length of the label sequence in different modeling units is quite different. In our corpus dataset, the maximum length of the label sequence in syllable initial/final with tone, syllable with tone and Chinese character are 119, 40, 40, respectively. The lengths of the label sequence in syllable with tone and Chinese character are the same, which are about one third of the label length in syllable initial/final with tone. Due to the different capabilities of DNN model architectures in classification, sequence feature extraction, and alignment between the acoustic features and the label tokens, the modeling units of transformer transducer affect the performance significantly.

The aim of our work is to research on the effect of different modeling units for transformer transducer. We compare the performance of transformer transducer with different modeling units and choose the best one for Mandarin speech recognition applications.

# 4. Results and Discussions

## 4.1. Dataset

All experiments are conducted on about 12,000 hours of Mandarin speech data with sampling rate in 8kHz and 16kHz, covering internal speech dataset in different business scopes of JD Digits and public speech dataset. The dataset are shown in Table 2. The maximum duration of the audio samples is 10s. The number of class and maximum length of the label sequence for different modeling units are shown in Table 3.

Table 2: *The dataset in our work.*

| Scope | Dataset | Sampling Rate | Duration |
|---|---|---|---|
| Internal | JD Digits dataset | 8kHz | ~5000h |
|  |  | 16kHz | ~5000h |
| public | AISHELL-1 | 16kHz | 178h |
|  | AISHELL-2 | 16kHz | 1000h |
|  | THCHS-30 | 16kHz | 30h |
|  | Primewords [24] | 16kHz | 100h |
|  | ST-CMDS [25] | 16kHz | 122h |
|  |  | **Total:** | ~12000h |

Table 3: *Dataset with different modeling units.*

| Modeling Units | Number of Class | Maximum Length of Label Sequence |
|---|---|---|
| syllable initial/final with tone | 227 | 119 |
| syllable with tone | 1256 | 40 |
| Chinese character | 7228 | 40 |

## 4.2. Experimental Setup

For all experiments, we use 80-dimension mix-bandwidth features computed on 20ms window with 10ms shift as shown in section 2.2. The acoustic feature encoding is composed of 3 convolutional layers, with filter dimension and channels (41,11,31), (21,11,32), (21,11,96) and the strides is (2,2), (2,1), (2,1) [26], which are followed by 10 self-attention blocks with 8 multi-head and 256 dimension. For the truncated transformer, we find that right context and left context have significant impact on the performance. Our experiments show that left = 20 (800ms) and right = 5 (200ms) obtain the best performance with a balance of accuracy and efficiency. The label feature encoding is composed with a 128-dimension embedding layer and two LSTM layers with 700 units [13]. The dimension of the joint net is 256 and the output dimension of the full connection layers is 228, 1257, 7229 for the three different modeling units.

All of our models are first pre-trained for the acoustic feature encoding using a CTC loss [27], then the model is fine-turned for transformer transducer. Our proposed model is optimized in the same method as [28]. And we set warm up steps to 8000 and the parameter $\lambda$ of learning rate to 0.5. The experiments are performed with a mini batch of 128 on 8 NVIDIA V100 GPUs.

## 4.3. Results

Figure 3 shows the fine-turned model learning curves with different modeling units. We can qualitatively compare the convergence of the model under different modeling units from the decline process of the loss function on the validation dataset. Results show that the best modeling units for model convergence is syllable with tone, when compared with syllable initial/final with tone and Chinese character.

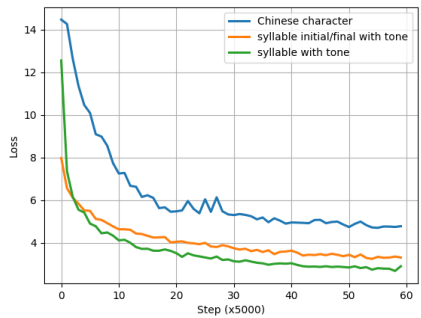

Figure 3: *Comparison of learning curves with various modeling units.*

We test the proposed method with various modeling units on the test set. To obtain a fair comparison, we use tokens and elements of syllable with tone as word units for WER and character units for Character Error Rate (CER), respectively. Thus, CER for Chinese character is not calculated. As shown in Table 4, the Mandarin transformer transducer using syllable with tone achieves the best performance. It yields an average of 14.4% and 44.1% relative WER reduction when compared with the models based on syllable initial/final with tone and Chinese character, respectively. It also outperforms the model based on syllable initial/final with tone with an average of 13.5% relative CER reduction.

Table 4: *Results on different test dataset.*

| WER/CER(%) |  | Modeling Units | | |
|---|---|---|---|---|
|  |  | syllable initial/final with tone | syllable with tone | Chinese character |
| **Dataset** | JD Digits dataset 1 | 7.51/4.76 | **7.06/4.31** | 7.77/NA |
|  | JD Digits dataset 2 | 13.01/8.52 | **12.03/8.07** | 14.46/NA |
|  | AISHELL-1 | 5.94/2.36 | **4.46/1.86** | 8.73/NA |
|  | AISHELL-2 | 7.24/2.94 | **6.14/2.64** | 11.31/NA |
|  | THCHS-30 | 16.28/10.04 | **13.66/8.26** | 31.80/NA |
|  | Primewords | 15.18/8.90 | **12.63/7.29** | 25.52/NA |
|  | ST-CMDS | 8.61/3.84 | **7.19/3.37** | 13.38/NA |
| **Average** |  | 10.54/5.91 | 9.02/5.11 | 16.14/NA |

# 5. Conclusions

A novel transformer transducer is proposed for Mandarin speech recognition and different modeling units are compared. In addition, a mix-bandwidth training method is proposed to improve the generalization performance of our model in real applications. Experimental results show that the syllable with tone achieves the best performance than syllable initial/final with tone and Chinese character. Since modeling units have very different effects on various models, each new ASR model needs to be compared with different modeling units, causing high computing cost. Our future work is to use automatic machine learning to automatically select the optimal modeling unit for a given ASR task.